# Results from the Worldwide Coma Morphology Campaign for Comet ISON (C/2012 S1)[1]

Nalin H. Samarasinha & Beatrice E.A. Mueller (Planetary Science Institute, USA), Matthew M. Knight (Lowell Obs, USA), Tony L. Farnham (Univ of Maryland, College Park, USA), John Briol (Spirit Marsh Obs, USA), Noah Brosch (Wise Obs, Tel Aviv Univ, Israel), John Caruso (Temecula, CA, USA), Xing Gao (No 1 Senior High School, Urumqi, China), Edward Gomez & Tim Lister (Las Cumbres Observatory Global Telescope Network, USA), Carl Hergenrother (Univ of Arizona, USA), Susan Hoban & Roy Prouty (Univ of Maryland, Baltimore County, USA), Mike Holloway (Holloway Comet Obs, USA), Nick Howes & Ernesto Guido (Remanzacco Obs, Italy), Man-To Hui (Univ of California, Los Angeles, USA), Joseph H. Jones, Tyler B. Penland, Samuel R. Thomas & Jim Wyrosdick (Univ of North Georgia, USA), Nikolai Kiselev & Aleksandra V. Ivanova (Main Astronomical Obs, National Academy of Sciences, Ukraine), Thomas G. Kaye (Raemor Vista Obs, USA), Jean-Baptist Kikwaya Eluo (Vatican Obs, USA), Betty P.S. Lau (Hong Kong), Zhong-Yi Lin (National Central Univ, Taiwan), José Luis Martin (Carpe Noctem Obs, Spain), Alexander S. Moskvitin (Special Astrophysical Obs, Russian Academy of Sciences, Russia), Martino Nicolini (Cavezzo Obs, Italy), Brian D. Ottum (Saline, MI, USA), Chris Pruzenski, David C. Vogel, Leo Kellett, Valerie Rapson, Joel Schmid, Brandon Doyle, Frank Dimino & Stephanie Carlino (Astronomy Section, Rochester Academy of Science, USA), Margarita Safonova, Jayant Murthy & Firoza Sutaria (Indian Institute of Astrophysics, India), David G. Schleicher (Lowell Obs, USA), Colin Snodgrass (The Open University, UK), Cihan T. Tezcan & Onur Yorukoglu (Ankara Univ, Kreiken Obs, Turkey), David Trowbridge (Tinyblue Obs, USA), Dennis Whitmer (Eagle Nest Digital Obs, USA), Quan-Zhi Ye (Univ of Western Ontario, Canada)

Corresponding Author information:
Nalin H. Samarasinha
Planetary Science Institute
1700 E Ft Lowell Road, Suite 106, Tucson, AZ 85719, USA
(Tel: 1-520-547-3952, Email: nalin@psi.edu)

[1] We dedicate this paper to our colleague and friend Martino Nicolini who on the 29th January 2015 passed away after a long fight against cancer. Martino, who was a nuclear engineer by profession, was an avid amateur astronomer and he collaborated with us as well as with many other organizations. He epitomized what the professional-amateur collaborations could accomplish and will be sorely missed by all.

# Abstract

We present the results of a global coma morphology campaign for comet C/2012 S1 (ISON), which was organized to involve both professional and amateur observers. In response to the campaign, many hundreds of images, from nearly two dozen groups were collected. Images were taken primarily in the continuum, which help to characterize the behavior of dust in the coma of comet ISON. The campaign received images from January 12 through November 22, 2013 (an interval over which the heliocentric distance decreased from 5.1 AU to 0.35 AU), allowing monitoring of the long-term evolution of coma morphology during comet ISON's pre-perihelion leg. Data were contributed by observers spread around the world, resulting in particularly good temporal coverage during November when comet ISON was brightest but its visibility was limited from any one location due to the small solar elongation. We analyze the northwestern sunward continuum coma feature observed in comet ISON during the first half of 2013, finding that it was likely present from at least February through May and did not show variations on diurnal time scales. From these images we constrain the grain velocities to ~10 m $s^{-1}$, and we find that the grains spent 2-4 weeks in the sunward side prior to merging with the dust tail. We present a rationale for the lack of continuum coma features from September until mid-November 2013, determining that if the feature from the first half of 2013 was present, it was likely too small to be clearly detected. We also analyze the continuum coma morphology observed subsequent to the November 12 outburst, and constrain the first appearance of new features in the continuum to later than November 13.99 UT.

# 1. Introduction

Comet C/2012 S1 (ISON) (hereafter comet ISON) was discovered on September 21, 2012 at a heliocentric distance, $r_h$, of 6.3 AU (Novski and Novichonok 2012). Soon it attracted worldwide interest because of its extremely small perihelion distance (0.0125 AU = 2.7 solar radii) and the prediction that it may become a bright naked eye object based on its brightness behavior soon after discovery. The discovery of comet ISON at a large heliocentric distance and the favorable observing geometry during most of the apparition, partially facilitated by the comparatively high orbital inclination (61.°9), made it the first instance that the behavior of cometary activity of a sungrazer[2] was monitored for a large range of heliocentric distances. This range included the large distances (i.e., > 2-3 AU) where the nuclear activity is dominated by super volatiles (e.g., CO, $CO_2$), the region where water is the primary driver of activity, as well as the extremely small heliocentric distances when the sublimation of refractory material, including metallic species that are present in the dust grains, could occur. In other words, the heliocentric distances covered a range over which the incident solar flux increased by a factor of about $2.5 \times 10^5$ making this the first time that the effects due to such a large range of solar fluxes on the nucleus were monitored for any comet. Anticipating these extreme conditions, there were many predictions on what might happen to the nucleus as it approached perihelion. The predictions included those pertaining to cometary activity and nuclear rotational state, as well as the ultimate fate of the nucleus (e.g., Samarasinha and Mueller 2013, Knight and Walsh 2013, Ferrin 2014).

Past work on morphological studies of cometary comae has demonstrated the value for inferring properties of the nucleus such as rotation period, seasonal activity changes, and pole orientation (e.g., Schleicher et al. 2003, Farnham et al. 2007, Farnham 2009, Knight et al. 2012 and references therein). With the goal of temporal monitoring of comet ISON's coma features, the campaign organizers[3] coordinated a Worldwide Campaign of Coma Morphology (see http://www.psi.edu/ison for the call for images). This effort was also carried out in coordination with the NASA Comet ISON Observing Campaign (CIOC) but was independent of the CIOC activities (see http://www.isoncampaign.org for additional information on CIOC). After May 2013, comet ISON was not observable for more than a few hours from any geographic location due to the relatively small solar elongation, so a global effort, incorporating images from both professionals and amateurs, was necessary to carry out a detailed analysis of the coma morphology and its temporal and spatial evolution. Although the relatively large geocentric distances to the comet during the apparition (>0.85 AU during the pre-perihelion leg and >0.43 AU during the post-perihelion leg) would not provide the ideal circumstances needed for producing high-resolution detailed coma features, the large range of heliocentric distances over which comet ISON might be studied and the resulting possibility of observing a variety of interesting coma morphologies motivated us to organize this coordinated worldwide campaign. The campaign solicited both continuum (dust) images as well as gas images of the near-nucleus region of the comet. We are happy to note that a large number of images were collected from observers spread around the world at different longitudes.

[2] A formal definition for the sungrazer comets is provided in Knight and Walsh (2013) in terms of the Roche limit.
[3] The campaign was organized by N.H. Samarasinha, B.E.A. Mueller, M.M. Knight, and T.L. Farnham.

Monitoring of comet ISON was encouraged for both before and after its perihelion passage, with an emphasis on the time around perigee (December 26, 2013) when it was expected to experience rapid changes and would have been observable for many hours per night in the northern hemisphere. However, the comet started to disintegrate just prior to its perihelion passage (cf. Knight and Battams 2014, Sekanina and Kracht 2014) leaving us with images only from the pre-perihelion leg.

In Section 2 of this paper, we provide a description of the images collected by the campaign. In Sections 3 and 4, we discuss the analyses of these images during the pre-water dominated phase and the water-dominated phase, respectively. Section 5 provides the summary and conclusions of this paper.

## 2. Observations

Images were collected from both amateur and professional observers, with diverse ranges of telescope apertures, observing conditions, filters, and data collection methodologies. These data are far from uniform, but such an approach was necessary in order to obtain as much longitudinal coverage as possible. Observers were asked to reduce (remove bias and correct for flat fielding) their own images prior to submitting them to the campaign. The campaign organizers, using these reduced images, carried out all image enhancements. The relevant enhancement techniques are described in Samarasinha and Larson (2014). The coma images we received cover a large interval from January 12, 2013 ($r_h$~5.14 AU) to November 22, 2013 ($r_h$~ 0.35 AU). Table 1 lists a summary of information on the images provided by different groups of observers ordered based on the geographic longitude of the observations. Figure 1 shows the observational circumstances for all the images received by the campaign. The symbols for different parameters listed in Table 1 and Figure 1 are identified in the legend of Figure 1. Most observations listed in Table 1 were not published elsewhere and are presented here for the first time. Observers submitted different sized images; however, the pixel scales for all the images were smaller than the respective astronomical seeing.

As can be seen from Table 1 and Figure 1, the majority of the images were obtained between September and mid-November, 2013, when the comet was bright and the solar elongation was >30˚. However, as we will discuss in detail in Section 4, the images from this time interval showed no clearly identifiable coma features in the continuum until the outburst that started around November 12 (e.g., Opitom et al. 2013a). In contrast, when the comet's solar elongation was favorable early in the apparition (i.e., before June, 2013), there were many clearly identifiable coma features in the continuum. In images of the gas coma, features appeared by November 1, nearly a month prior to the perihelion of November 28.779 UT (e.g., Opitom et al. 2013b, Knight and Schleicher 2015). In the following sections, we analyze the continuum images and provide a rationale for the morphological behavior.

Table 1. Summary of observations corresponding to different observers

| Observers | Longitude, Latitude | Range of UT dates | Filters | $r_h$ [AU] | Δ [AU] | $\alpha$ [deg] | $\varepsilon$ [deg] | PA⊙ [deg] |
|---|---|---|---|---|---|---|---|---|
| Z.-Y. Lin, H.-Y. Hsiao, C.-S. Lin, H.-C. Lin | 120.87° E, 23.47° N | Oct 4-6, 8-10, 12, 14, 17-24, 26-28 | R | 1.6-1.1 | 2.1-1.3 | 28-48 | 49-53 | 112-113 |
| | | Nov 5-8, 11, 13-14, 16, 22 | R | 0.9-0.4 | 1.1-0.9 | 57-101 | 50-21 | 114-101 |
| Q.-Z. Ye, M.-T. Hui, X. Gao | 87.01° E, 43.05° N | Oct 14, 20-21, 26, 31 | open | 1.4-1.0 | 1.8-1.2 | 34-51 | 52-52 | 112-113 |
| | | Nov 1, 6, 13-16 | open | 1.0-0.6 | 1.2-0.9 | 52-83 | 52-35 | 114-110 |
| M. Safonova, J. Murthy, F. Sutaria, N. Brosch | 78.96° E, 32.78° N | Jan 22 | V,R,I | 5.0 | 4.1 | 3 | 164 | 315 |
| | | Feb 19, 21-22 | R, I | 4.7-4.7 | 4.0-4.0 | 9-10 | 132-128 | 284-283 |
| | | May 1, 4 | V,R, open | 3.9-3.9 | 4.3-4.3 | 13-13 | 59-56 | 270-270 |
| | | Sep 8 | I | 2.1 | 2.8 | 17 | 36 | 111 |
| | | Oct 1, 3 | B,R | 1.7-1.6 | 2.2-2.1 | 27-28 | 48-48 | 111-112 |
| | | Nov 10 | R | 0.8 | 1.0 | 67 | 45 | 113 |
| N. Kiselev, A. Ivanova, A. Moskvitin | 41.44° E, 43.65° N | Nov 11 | B, V, R, I | 0.7 | 1.0 | 69 | 44 | 113 |
| C. Tezcan, O. Yorukoglu | 32.77° E, 39.84° N | Oct 27 | V, R, I | 1.1 | 1.4 | 46 | 53 | 113 |
| P.S. Lau | 2.30° W, 38.15° N | Sep 28, 30 | R | 1.7-1.7 | 2.2-2.2 | 25-26 | 46-48 | 111-111 |
| | | Oct 2, 19, 25, 31 | R | 1.6-1.0 | 2.1-1.2 | 27-51 | 48-52 | 112-114 |
| | | Nov 1, 6, 14 | R | 1.0-0.6 | 1.2-0.9 | 52-78 | 52-38 | 114-112 |
| J. Martin | 3.68° W, 40.40° N | Sep 24 | V, R | 1.8 | 2.4 | 23 | 44 | 111 |
| | | Oct 6-8, 11-12, 16, 31 | V, R, open | 1.6-1.0 | 2.0-1.2 | 29-51 | 50-52 | 112-114 |
| | | Nov 12, 14 | V, R | 0.7-0.6 | 1.0-0.9 | 71-78 | 43-38 | 113-112 |
| J. Caruso | 16.51° W, 28.30° N | Oct 6 | R | 1.6 | 2.0 | 29 | 50 | 112 |
| N. Howes, E. Guido, M. Nicolini | 17.88° W, 28.76° N | May 1-3, 5-7 | R | 3.9-3.8 | 4.3-4.3 | 13-12 | 59-53 | 270-270 |
| | | Oct 1, 4-11,17,20-23 | R | 1.7-1.2 | 2.2-1.5 | 27-42 | 48-54 | 111-113 |
| | | Nov 4, 6-10, 12-13 | R | 0.9-0.7 | 1.1-0.9 | 56-74 | 50-41 | 114-112 |
| S. Hoban, R. Prouty | 76.71° W, 39.25° N | Nov 4, 8-10 | CN, GC, C2, H2O+, open | 0.9-0.7 | 1.1-1.0 | 56-68 | 50-44 | 114-113 |
| D. Vogel | 77.40° W, 42.93° N | Nov 14, 16 | open | 0.7-0.6 | 0.9-0.9 | 77-78 | 40-38 | 112-111 |
| C. Pruzenski, L. Kellett, V. Rapson, J. Schmid, B. Doyle, F. Dimino, D. Vogel, S. Carlino | 77.50° W, 42.93° N | Nov 8, 11, 14, 16 | open | 0.8-0.6 | 1.0-0.9 | 63-83 | 47-35 | 114-110 |
| B. Ottum | 83.71° W, 42.19° N | Oct 8, 18, 25, 30 | open | 1.5-1.0 | 1.9-1.3 | 31-50 | 50-52 | 112-114 |
| | | Nov 8, 12 | open | 0.8-0.7 | 1.0-0.9 | 63-73 | 47-42 | 114-112 |
| J. Jones, T. Penland, J. Wyrosdick, S. Thomas | 84.05° W, 34.52° N | Nov 2, 14 | V | 1.0-0.6 | 1.2-0.9 | 53-78 | 52-38 | 114-111 |

| Observers | Longitude, Latitude | Range of UT dates | Filters | $r_h$ [AU] | Δ [AU] | $\alpha$ [deg] | $\varepsilon$ [deg] | PA⊙ [deg] |
|---|---|---|---|---|---|---|---|---|
| M. Holloway | 94.40° W, 35.54° N | Sep 3, 6, 26 | open | 2.1-1.7 | 2.9-2.3 | 15-25 | 33-46 | 111-111 |
| | | Oct 7, 11, 24-25 | open | 1.5-1.1 | 2.0-1.4 | 30-44 | 50-53 | 112-113 |
| | | Nov 1, 7 | open | 1.0-0.8 | 1.2-1.1 | 52-62 | 52-48 | 114-114 |
| J. Briol | 94.89° W, 45.66° N | Oct 13, 18, 25, 27 | open | 1.4-1.1 | 1.8-1.4 | 34-46 | 52-53 | 112-114 |
| | | Nov 3, 12, 15 | open | 0.9-0.6 | 1.2-0.9 | 54-80 | 51-37 | 114-111 |
| E. Gomez, Z-Y. Lin, T. Lister | 104.02° W, 30.68° N | Nov 2, 10-11,15,17 | B, V, I | 1.0-0.6 | 1.2-0.9 | 53-84 | 52-34 | 114-110 |
| J-B. Kikwaya Eluo | 109.89° W, 32.70° N | Sep 28-30 | B, V, R | 1.7-1.7 | 2.2-2.2 | 25-26 | 46-48 | 111-111 |
| | | Nov 6-8 | B, V, R | 0.9-0.8 | 1.1-1.0 | 59-64 | 49-47 | 114-114 |
| C. Hergenrother | 109.89° W, 32.70° N | Jan 15, 20-21 | V, R | 5.1-5.0 | 4.1-4.1 | 2-3 | 170-165 | 346-316 |
| | | Feb 8, 15 | V, R | 4.9-4.8 | 4.0-4.0 | 7-8 | 145-136 | 290-286 |
| | | Mar 16-17 | V, R | 4.4-4.4 | 4.1-4.1 | 13-13 | 104-101 | 276-276 |
| | | Apr 21-22 | V, R | 4.0-4.0 | 4.3-4.3 | 13-13 | 68-66 | 271-271 |
| | | May 19-21 | V, R | 3.7-3.6 | 4.3-4.3 | 11-11 | 43-41 | 269-269 |
| B. Gary | 110.23° W, 31.45° N | Oct 8, 31 | R, open | 1.5-1.0 | 1.9-1.3 | 31-50 | 50-52 | 112-114 |
| D. Whitmer, B. Gary | 110.25° W, 31.42° N | Oct 16, 31 | R | 1.4-1.0 | 1.7-1.3 | 34-50 | 53-52 | 113-114 |
| | | Nov 1-2, 4, 6, 9 | V, R | 1.0-0.8 | 1.2-1.0 | 52-66 | 52-45 | 114-113 |
| T. Kaye, B. Gary | 110.65° W, 31.67° N | Oct 20, 28 | V, R | 1.3-1.1 | 1.6-1.3 | 39-48 | 53-53 | 113-114 |
| | | Nov 2 | R | 1.0 | 1.2 | 53 | 52 | 114 |
| M. Knight, D. Schleicher | 111.42° W, 34.74° N | Jan 12, 13 | R | 5.1-5.1 | 4.2-4.2 | 2-2 | 170-170 | 355-8 |
| | | Mar 11 | B, V, R | 4.5 | 4.1 | 12 | 109 | 277 |
| | | Apr 4, 6, 7, 8 | B, V, R, I, CN, BC | 4.2-4.2 | 4.2-4.2 | 14-14 | 84-80 | 273-272 |
| | | May 1, 3 | B, V, R, I | 3.9-3.9 | 4.3-4.3 | 13-13 | 59-56 | 270-270 |
| | | June 11 | V, R, I | 3.4 | 4.3 | 7 | 25 | 265 |
| | | Sep 12-13, 30 | V, R, OH, CN, C3, CO+, BC, C2, GC, RC | 2.0-1.7 | 2.7-2.2 | 18-26 | 38-48 | 111-111 |
| | | Oct 1-6, 8-9, 15 | B, V, R, I, CN, C3, CO+, BC, C2, GC, RC | 1.7-1.4 | 2.2-1.7 | 27-36 | 48-53 | 111-113 |
| | | Nov 1-4, 6-12 | R, CN | 1.0-0.7 | 1.2-1.0 | 52-72 | 52-42 | 114-113 |
| D. Trowbridge | 122.60° W, 48.08° N | Oct 6, 29 | V, R | 1.6-1.1 | 2.0-1.3 | 29-49 | 50-53 | 112-114 |
| T. Lister, C. Snodgrass, M. Knight | 156.26° W, 20.71° N | Oct 22-25 | G, R, I | 1.2-1.2 | 1.5-1.4 | 41-44 | 54-54 | 113-113 |
| | | Nov 3-4, 8-9 | R, OH, CN, C3, C2, NH2 | 0.9-0.8 | 1.2-1.0 | 54-65 | 51-46 | 114-113 |

Notes:

1. The symbols for the parameters are defined in Figure 1.
2. "open" filter refers to either "no filter" or "clear filter".
3. B, V, G, R, and I are broadband filters (could be of different photometric systems) while all others are narrowband filters.
4. Not all filters were used during every night of a given month.

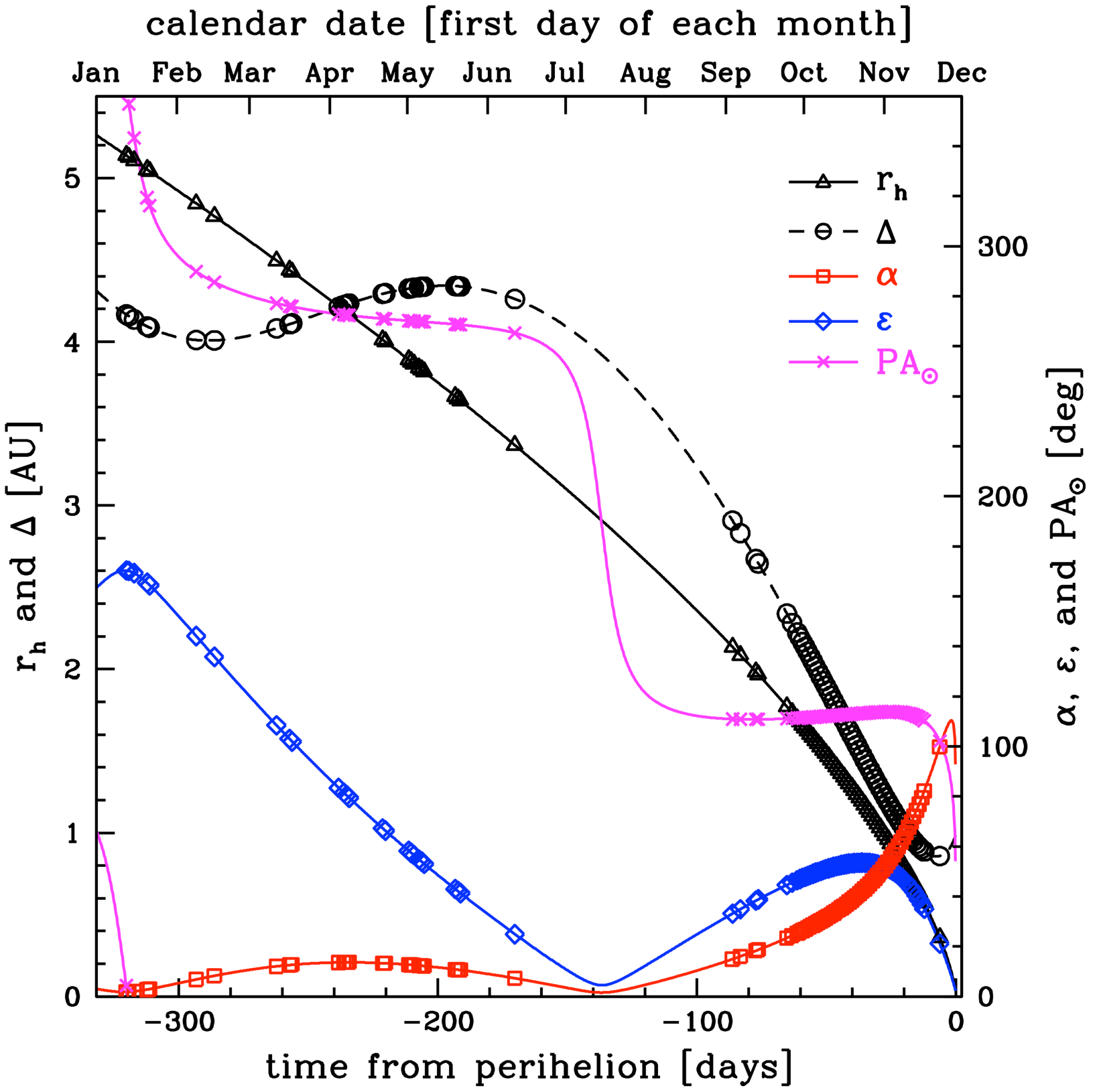

Figure 1. Behavior of heliocentric distance $r_h$, geocentric distance $\Delta$, solar phase angle $\alpha$, solar elongation (Sun-Earth-comet angle) $\varepsilon$, and position angle PA of the skyplane-projected Sun direction $PA_{\odot}$ (measured from north through east) as a function of time. The symbols depict the UT dates when observations are available from this global campaign. The parameters corresponding to each symbol are identified in the legend.

## 3. Results from the Pre-Water-Dominated Phase

The pre-water-dominated phase (i.e., when the surface temperature of a comet is too low for water to be the dominant or the most productive volatile, where super-volatiles such as CO or $CO_2$ could be the primary driver of cometary activity) occurs when a comet is further than about 2-3 AU from the Sun (e.g., Bockelée-Morvan et al. 2004). The time when comet ISON was at small solar elongations (i.e., approximately < 30° from June through August 2013; also Figure 1) coincides with the transition to water-dominated gas production, providing a natural boundary for the observations. Therefore, we designate observations prior to July 2013 (i.e., $r_h$>~3.1 AU) as representing the pre-water-dominated phase, while images after August 2013 (i.e., $r_h$<~2.2 AU) as coinciding with the water-dominated phase. In this section, we analyze and discuss the coma morphology and its evolution in the pre-water dominated phase.

Despite the fact that the comet was discovered at 6.3 AU from the Sun, the large geocentric distance coupled with the low surface brightness of the coma prevented the acquisition of high signal-to-noise (S/N) images with good spatial resolution. In addition, viewing of the impact of radiation pressure is the combined effect of the large heliocentric distance and the small solar phase angle (<10° until late-February with a minimum of 1.8° on January 11). This impeded any immediate detection of unambiguous coma morphology other than the dust tail in the earliest images (Figure 2). However, as the comet moved towards the inner solar system it underwent a rapid increase in its brightness prior to 5 AU (e.g., Figure 2 of Meech et al. 2013) that provided the first realistic opportunity for coma morphological studies.

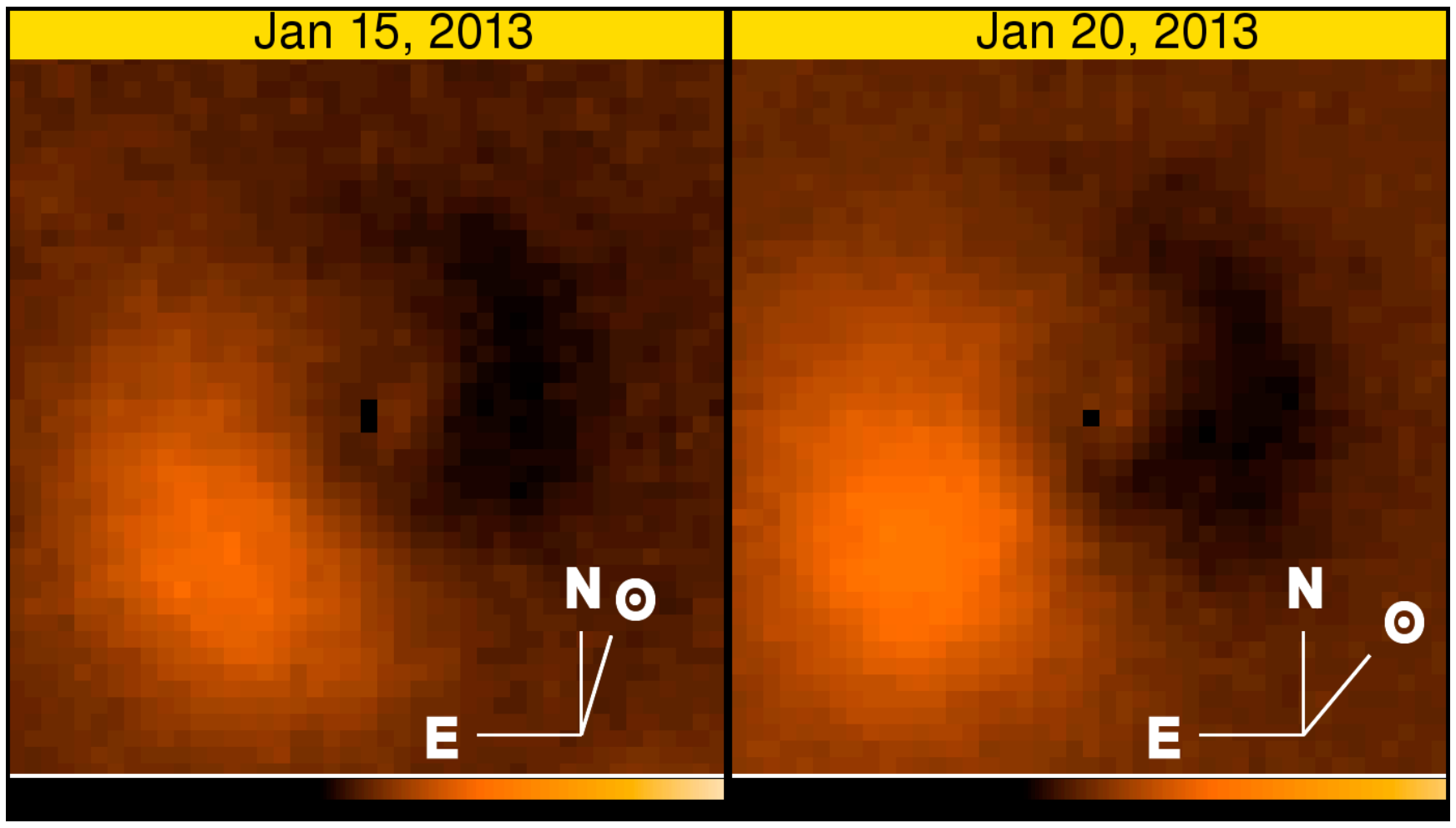

Figure 2. Inner coma region of two R-band images taken at the Vatican Advanced Technology Telescope in mid-January ($r_h$~5 AU). The images are enhanced using division by an azimuthal median technique (e.g., Samarasinha and Larson 2014). Orange represents brighter regions while black denotes the dimmer regions. Each panel is approximately 48,000 km across with the nucleus located at the center. The skyplane-projected PA of the Sun direction was rapidly moving clockwise from 343° to 320° (respective values are for January 15 and 20). The bright feature in the southeast quadrant is attributed to dust emitted from the nucleus that was subsequently swept away in the respective anti-sunward directions over a few weeks. Low spatial resolution and S/N due to the large geocentric distance prevent unambiguous detection of coma features close to the nucleus.

## 3.1 Sunward Feature in the Continuum Images

The first announcement of the presence of a coma feature in the continuum was based on enhanced Hubble Space Telescope (HST) images taken on April 10, 2013 (Li et al. 2013). The HST images showed a northwest sunward feature starting at a westerly direction and curving towards north and then merging with the tail due to radiation pressure. Then, Howes and Guido (2013) announced the presence of the same sunward feature in ground-based images taken in early-May ($r_h$~3.9 AU) from the 2m Liverpool Telescope (LT). Figure 3 shows the entire set of the LT images from this epoch indicating that there are no clearly discernible changes in the morphology on a daily timescale. Image enhancement of comet ISON images taken at the 4.3m Discovery Channel Telescope (DCT) by Knight and Schleicher (2015) provides clear evidence that the same sunward feature was present from March through mid-May, and HST observations by Hines et al. (2014) in early May confirm that it was relatively unchanged since April even at higher spatial scales. Images of comet ISON taken at the 1.8m Vatican Advanced Technology Telescope (VATT) by Carl Hergenrother in mid-February show the same sunward feature[4] albeit at a lower S/N. Therefore, we infer that this sunward feature is present at least in the images spanning an interval from mid-February to mid-May 2013. As June approached, the decreasing solar elongation prevented continued monitoring of this feature[5]. Figure 4 shows the temporal evolution of the feature in the February-May time frame.

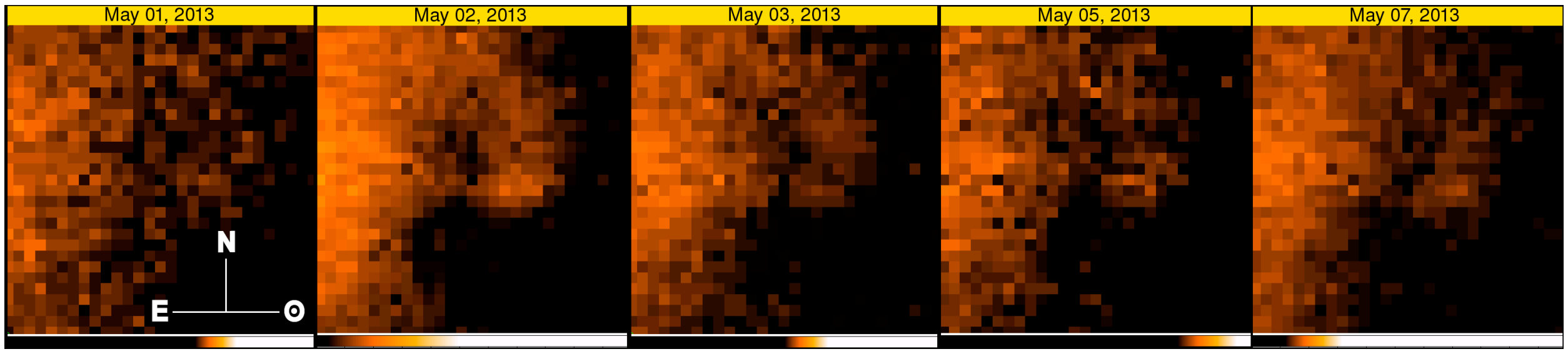

Figure 3. The near nucleus region of R-band images taken at the Liverpool Telescope on May 1, 2, 3, 5, and 7 (left to right). Images are enhanced using the division by azimuthal average technique, which is nearly identical to the division by azimuthal median (e.g., Samarasinha and Larson 2014). Each panel is approximately 25,000 km across with the nucleus located at the center. The skyplane-projected PA of the Sun direction is 270°. The sunward feature originates in an approximately westerly direction (to the right) and then moves northward and finally merges with the dust tail which is to the left (East) as a consequence of the radiation pressure effects. No clearly discernable temporal variations of the sunward feature are detected at this spatial resolution and scale. The minor changes from one panel to another are due to low brightness of the comet at this heliocentric distance.

For activity originating from a fixed source region away from the poles, one expects to see morphological variations on rotational timescales. No photometric variations in the brightness of comet ISON were detected in January by the Deep Impact spacecraft (Farnham et al. 2013), by the HST in April (Li et al. 2013), or by the Spitzer Space Telescope in June (Lisse et al. 2013) suggesting any rotational variation should be small. Despite that, observations taken on November 1, 2013 by the HST, when the comet was much closer to us ($r_h$=1.00 AU, $\Delta$=1.23 AU), show a single-peak photometric variation of nearly a factor two with a period of ~10.4 hours which was attributed to the rotation of the nucleus (Lamy et al. 2014).

[4] Generally, we are extremely reluctant to trust the reality of the sunward feature present in these mid-February images (and to an extent in some March images), as the spatial location of the feature is extremely sensitive to a single pixel offset of the nucleus position. However, the fact that the PA of the feature and the general morphology are consistent with subsequent April through May images provide confidence that the feature is indeed real.

[5] Knight and Schleicher (2015) have images from June 11; however, these images were taken at extremely high airmass and have S/N too low to detect coma features.

However, no variations in the morphology were detected at rotational timescales either based on the ground-based images or on the HST images. Due to this fact, Li et al. (2013) attribute the origin of the sunward feature observed by HST in April to a source region near comet ISON's rotational pole. Another possibility is that this feature is due to the cumulative grain outflow from the sunward side of the nucleus as a response to insolation and not necessarily from a fixed source region on the nucleus (cf. Belton 2013).

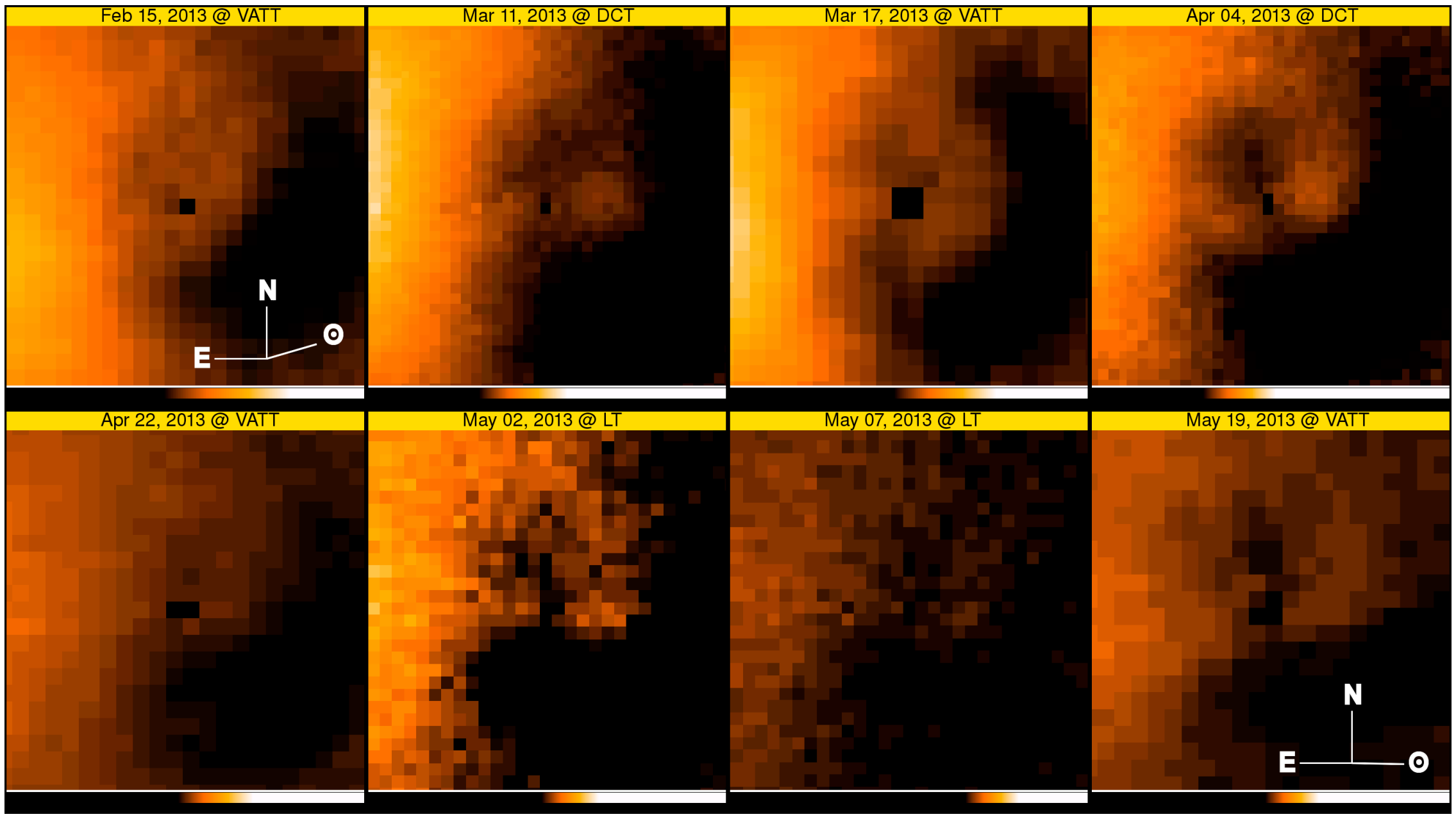

Figure 4. Images from February to May ($r_h$ decreasing from ~4.8 AU to ~3.7 AU) showing the evolution of the sunward feature. The date and telescope are given at the top of each panel, and the corresponding geometric circumstances can be determined from Figure 1 and Table 1. All images are enhanced with division by azimuthal median. Each panel is approximately 25,000 km across. The PA of the Sun direction gradually varies from 286° (February 15) to 269° (May 19). The sunward feature in the northwest quadrant was morphologically very similar during the entire time frame.

## 3.2 Characterization of Grains in the Sunward Feature

The measured sunward extension of the feature projected onto the skyplane, $d$, is $4{\times}10^3$ – $6{\times}10^3$ km. Therefore, for this $d$, $V^2/\beta > 10^{-3}$ km$^2$ s$^{-2}$ where $V$ is the outflow velocity of grains and $\beta$ is the radiation pressure parameter (cf. equation (5) of Mueller et al. 2013). For dominant micron-size grains (i.e., $\beta \sim 0.1$; Burns et al. 1979), $V$ should be ~10 m s$^{-1}$. Grain outflow velocities for micron sized grains that are of the order of ten or a few tens of m $s^{-1}$ are not uncommon and were observed in other comets (e.g., comet Siding Spring (C/2013 A1): Li et al. (2014), Tricarico et al. (2014), Kelley et al. (2014); comet 9P/Tempel 1: Meech et al. (2011), Vasundhara (2009)).

Considering that the corresponding geocentric distance during this interval is approximately 4.0 – 4.3 AU and assuming a typical ground-based astronomical seeing of 1 arcsec, the spatial resolution at the comet is ~3x$10^3$ km (for HST, the corresponding spatial resolution is an order of magnitude smaller). During a time interval corresponding to the single peak rotational period near 10.4 hours suggested by Lamy et al. (2014), grains would have moved of the order of $4\times10^2\times\cos\gamma$ km in the skyplane where $\gamma$ is the angle between the skyplane and the initial direction of the feature. This distance is an order of magnitude smaller than the ground-based seeing disk, so it is not surprising that we do not see any fine structure of the sunward feature in the ground-based images. The angular resolution of the HST images is comparable to the spatial displacement of grains in the sunward feature over a diurnal cycle. Therefore, despite the better angular resolution, even for HST images, we do not have sufficient spatial resolution to detect any diurnal scale variations in the coma structure.

On the other hand, the long-term variations over timescales of months in the sunward feature during the mid-February to mid-May interval are generally consistent with the evolution of the Earth-comet-Sun geometry and in particular the projected solar direction with respect to the comet in the skyplane (Figures 1 and 5). Based on the derived grain outflow velocity and the fact that we can observe the ultimate merging of the sunward feature with the tail due to radiation pressure, we estimate that by the time the grains reach the tail, they must have spent 2-4 weeks in the coma after being ejected from the nucleus (cf. equation (7) of Mueller et al. 2013).

## 3.3 Dynamical Constraints Based on the Sunward Feature

In general, if a coma feature originates at or near a pole, the pole direction should lie in a half-great circle defined by the PA of that feature. When the Earth direction as seen from the comet changes due to orbital motions, the PA of the feature may also change. One can use a range of PAs corresponding to observations made at different times to determine a pole solution by considering the intersection of these half-great circles.

In the case of comet ISON, the Earth direction changed less than 8° from mid-February to mid-May (cf. Figure 5). In addition, the lack of good spatial resolution of the sunward feature due to the comparatively large geocentric distances to the comet resulted in large error bars for the PAs for the sunward feature (cf. Figure 4). Therefore, from these observations, the determination of a robust pole direction based on intersections of multiple half-great circles corresponding to respective PAs is not feasible. In this case, most half-great circles appear to intersect near the Earth's direction, which are confined close to each other. This result is simply a manifestation of the inadequacy of this technique to yield reliable results when the Earth direction has not changed significantly with time, and our data provide no additional constraints to the solution defined by Li et al. (2013) in their Figure 4. This result was based on the higher resolution HST observations (see Figure 5) with the assumption that the feature originated from a fixed source region at or near the pole. As commented earlier in Section 3.1 (and also as favored by Knight and Schleicher 2015), the feature could simply be the cumulative dust outflow from the sunward side (with a significant contribution coming from regions near the sub-solar point) and in that case the sunward feature cannot be used to constrain the pole. Detailed modeling of this scenario is beyond the scope of this paper.

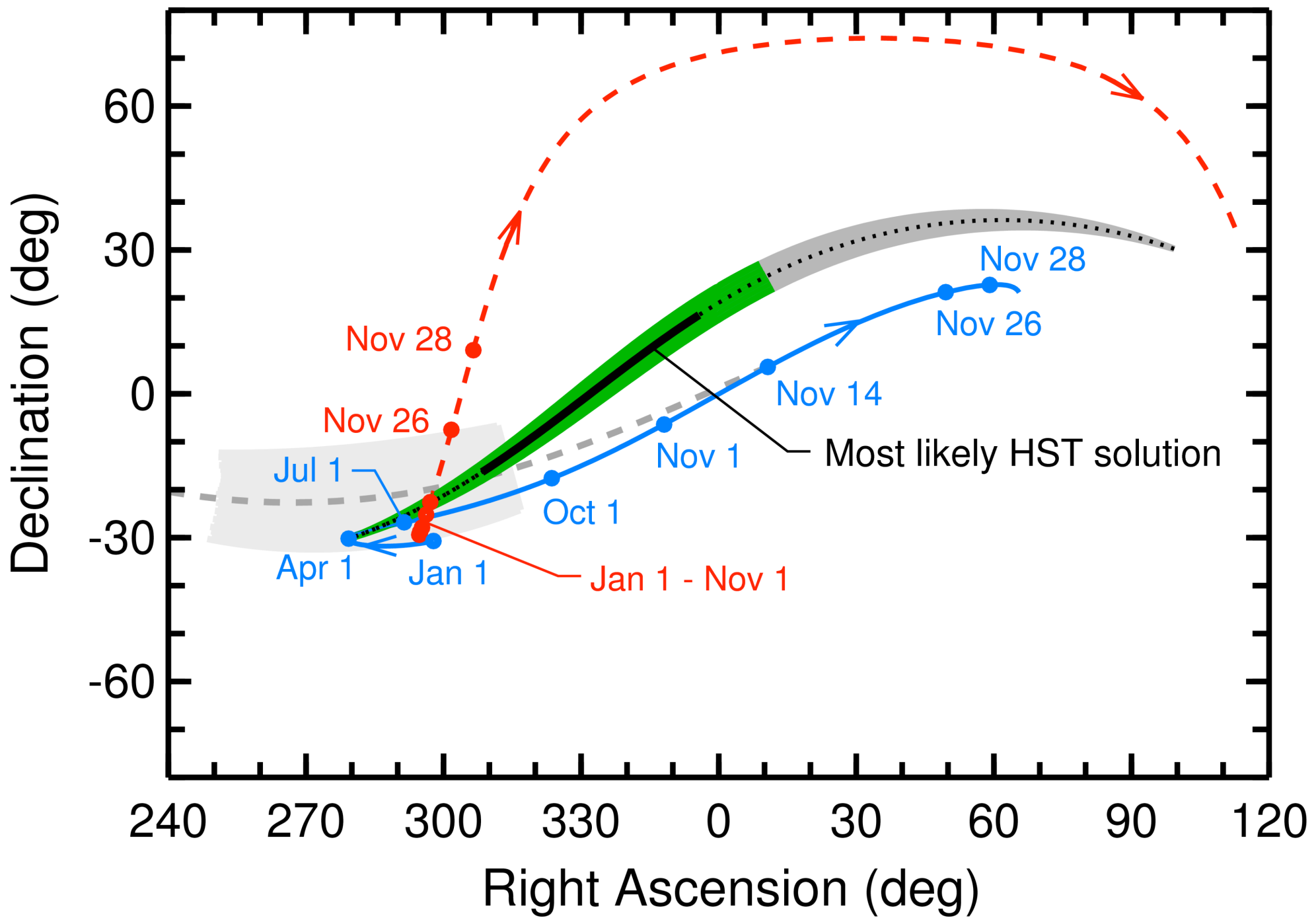

Figure 5. The family of pole solutions based on Li et al. (2013) is shown by the half-great circle (with its width corresponding to the error in the determination of the PAs). The Earth and Sun directions as seen from the comet from January 1, 2013 up to perihelion are represented by blue solid and red dashed lines, respectively. The dots denote the respective directions at 0 UT on January 1, April 1, July 1, October 1, November 1, November 26, and November 28 while the direction of arrows denote the change in directions with time. The dots for the Sun direction partially overlap from January 1 through November 1 indicating that the Sun motion is minimal during this time; the huge change in Sun direction from November 28 (0 UT) until perihelion corresponds to <19 hrs. The green and dark gray parts of the half-great circle denote the portions in sunlight and darkness respectively during the HST observations. The Earth direction during the HST observations is nearly the same as that for April 1. The rightmost end of the solid black line (which depicts the "most likely HST solution") represents the pole solution that is in the skyplane. The dashed grey line represents the great circle solution from the November outburst observation (Section 4.3) with the light grey envelope on the left indicating the range for the pole solution for a 10° uncertainty in the PA and 30° uncertainty with respect to the skyplane. For all the pole solutions in the figure, the diametrically opposite solutions are also valid but are not shown.

# 4. Results during the Water-Dominated Phase

The coma campaign during the water-dominated phase focused on both continuum as well as gas images. However, there were only a few observers who had access to narrowband gas filters: Hoban and Prouty (Telescope at University of Maryland, Baltimore County), Lister et al. (Faulkes Telescope North as part of the Las Cumbres Observatory Telescope Network), and Knight and Schleicher (using various telescopes at Lowell Observatory). The gas images of Knight and Schleicher were the only images received by the campaign that had sufficient S/N to detect unambiguous coma features. As these images are already discussed in detail in Knight and Schleicher (2015), we will not consider them further. However, we point out that unambiguous gas coma features were seen starting nearly a month prior to the perihelion (cf. Opitom et al. 2013b, Knight and Schleicher 2015). In the remainder of this section, we will concentrate on the behavior of the continuum features.

## 4.1 Searching for Coma Features in the Continuum

As September 2013 approached, the solar elongation increased sufficiently (to ~30°) for ground-based observations despite the observing window on any given night being short and the comet being at high airmass. Observations from September through November were also facilitated by the decreasing heliocentric and geocentric distances (Figure 1) and the resultant increase in the comet's apparent brightness (e.g., Meech et al. 2013). However, up until about two weeks prior to the perihelion (i.e., $r_h$~0.65 AU), we could not detect any unambiguous evidence of continuum features, except for the dust tail, even with image enhancements. To illustrate that an unambiguous continuum feature was not detected, we show in Figure 6, a high S/N image from October 17 ($r_h$~1.3 AU) representative of the middle of the observing window when water was the dominant volatile. This figure demonstrates the dependency of the "sunward feature" on the assumed nucleus (center) location. As shown elsewhere using numerically simulated images (Figure 9 of Samarasinha and Larson 2014), in the absence of any corroborating evidence, one should be extremely cautious to trust the reality of an apparent coma structure after image enhancement if that structure is sensitive to minute changes in the chosen nucleus location such as a one-pixel offset. We carried out a similar test for the September 12 images taken at the DCT by Knight and Schleicher (2015) and in that case the feature is more stable even though the PA of the feature is still sensitive to the chosen nucleus location. We attribute this behavior to the relatively high S/N as well as the smaller pixel size of the DCT images. Even if the sunward feature is real in early-September, we assert that its skyplane extent must be much smaller than in the February to May time frame. I.e., at best, it can only be marginally detectable.

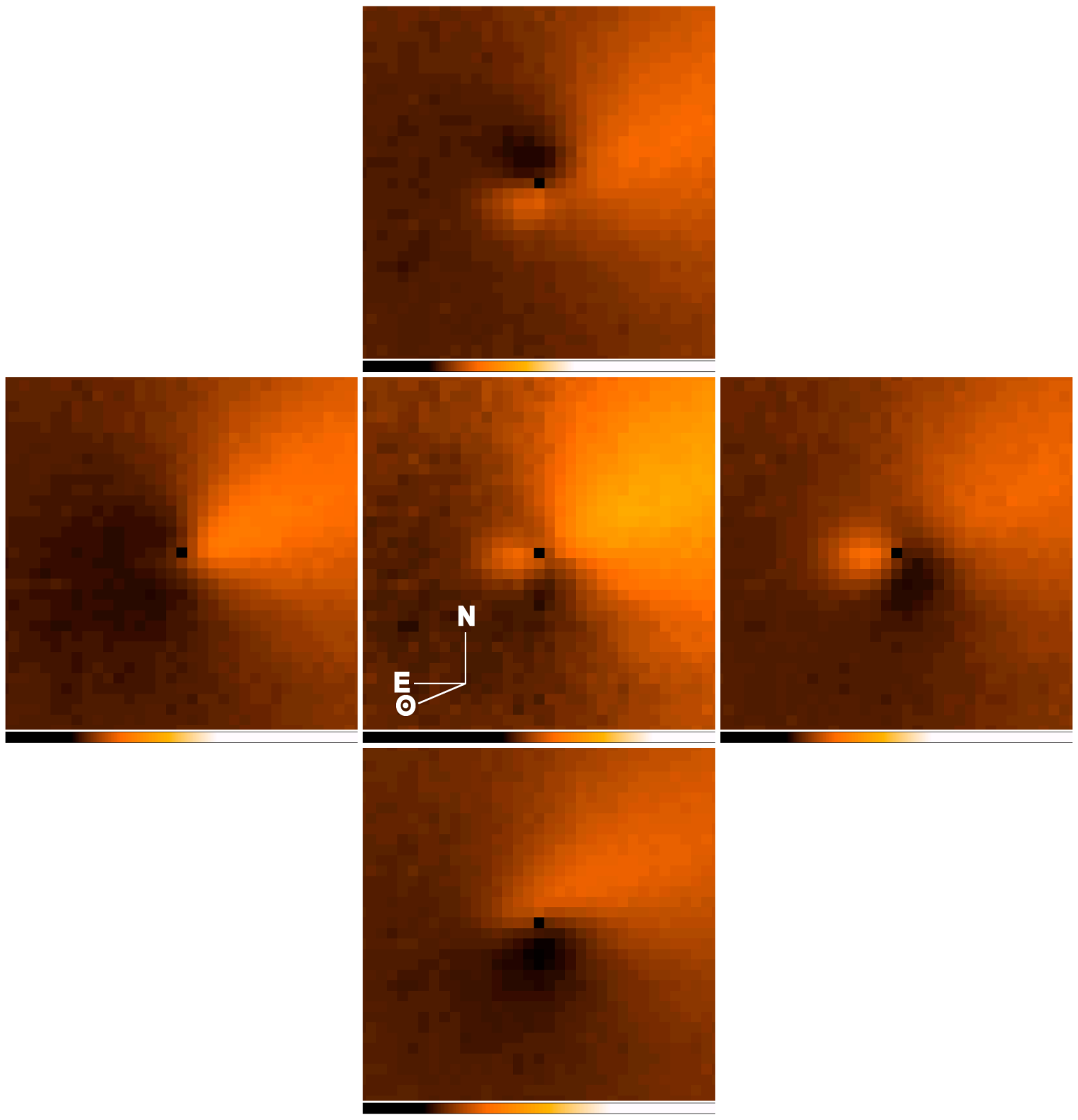

Figure 6. A high S/N continuum image of comet ISON taken on October 17 ($r_h$~1.3 AU) at the Liverpool Telescope. The image is enhanced using the division by azimuthal average technique. In the center panel, the optocenter is taken to represent the nucleus. In the other panels, the center is shifted by one pixel in the respective directions. Each panel is approximately 12,000 km across. The Sun is at a PA of 113°. The dust tail is at a westerly/northwesterly direction. Note that the apparent sunward feature in the center panel cannot be seen in all the panels at the same orientation.

## 4.2 Why Were no Features Detected in the Continuum?

What happened to the sunward feature observed during the pre-water-dominated phase? Did it disappear, as water became the dominant volatile responsible for cometary outgassing? Since the change in the Sun direction as seen from the comet from May to September is only about 3° (cf. Figure 5), it is unlikely that the lack of a feature was caused due to a change in the insolation geometry.

If one assumes that (a) the sunward feature during the pre-water-dominated phase retained the same character as during the water-dominated phase, (b) the change in the dust outflow velocity $V$ in the intervening period is small, and (c) the change in the projection effects can be ignored[6], then the sunward extent of the feature $d$ should decrease as $r_h^2/\sin\alpha$ (cf. equation (4) of Mueller et al. 2013). In late-September when $r_h$ is nearly half of that in mid-May and $\alpha$ has nearly doubled, *d* should be ~700 km. As $\Delta$ ~ 2.3 AU, the angular extent of the sunward feature should be ~0.4 arcsec which is smaller than the typical astronomical seeing. Therefore, it is not surprising that we cannot make a clear identification of the sunward feature in late-September even if it indeed existed. In October and November, $\alpha$ increases further while $r_h$ decreases, making the skyplane extent of the feature even smaller. HST images taken in the continuum on October 9 ($r_h$~1.5 AU) and November 1 ($r_h$~1.0 AU) do not show a sunward feature, or another feature for that matter, other than the dust tail (J.-Y. Li; personal communications, 2013). This suggests that, at least around the times when these HST images were taken, the skyplane extent of a sunward feature was extremely small (<100 km). Another possibility is that such a feature was absent.

[6] For many of the "most likely solutions" suggested for the jet/pole direction in Figure 4 of Li et al. (2013) (also Figure 5), the projection effects tend to make $d$ even smaller during the September to October time frame (compared with the February to May time frame) based on the corresponding Earth directions if the feature is caused by a fixed source region on the nucleus.

If the direction of the sunward feature was close to the skyplane during the mid-February to mid-May time frame, then during late-September, the feature could still maintain a PA approximately in the same general westerly to north-westerly direction as that during the pre-water dominated phase. To illustrate this, in Figure 5, we show the family of solutions suggested for the pole in Li et al. (2013) as well as the Sun and Earth directions as a function of time. However, as seen from Figure 1, the PA of the solar direction in the skyplane changed dramatically in July and by late-September the PA of the Sun was around 110°. That means the February-May sunward feature would be essentially in the tailward direction in late-September, thus effectively precluding its detection. We emphasize that this argument is relevant only if the feature is due to a fixed source region near the pole, the pole direction is close to the skyplane during the mid-February to mid-May interval, and this direction was nearly the same in September.

## 4.3 Behavior of the Coma in the Month Before Perihelion

A series of observations indicating rapid increases in activity and outburst events were reported in the month before perihelion. By November 5 ($r_h$~0.9 AU), water production rate increased by ~50% from two days earlier, and that of other gas species increased by twice the corresponding amount, but no change was observed in the dust production (Opitom et al. 2013b). On November 12 ($r_h$~0.7 AU), water production rate was ~50% higher than the previous night, and continued to increase by nearly an order of magnitude over the next two days. Dust production also increased by an order of magnitude by November 14 (Opitom et al. 2013a). Likely associated with this outburst, two arclet-like “wings” were detected (Boehnhardt et al. 2013) in the coma for the first time around November 14.2 UT, and confirmed on November 14.37 (C. Opitom; personal communications, 2015) and November 14.99 (Ye et al. 2013). These “wings”, which originate from the nucleus at PAs nearly perpendicular to the tail and curve back in the tailward direction, are likely to be associated with fragmentation[7] of the nucleus (e.g., Harris et al. 1997, Tozzi et al. 1997, Farnham et al. 2001, Jehin et al. 2002, Hadamcik and Levasseur-Regourd 2003, Boehnhardt 2004, Farnham 2009). A third increase in the gas production was reported (Opitom et al. 2013c) to have started between November 18 and 19 ($r_h$~0.5 AU). These observations reveal behavior that is remarkably similar to that of comet D/1999 S4 (LINEAR) before its complete breakup. One of the LINEAR outbursts was connected to a fragmentation event, with associated “wings” in the coma, that may have been related to the final disintegration of the nucleus (e.g., Farnham et al. 2001, Tozzi and Licandro 2002). The existence of “wings” in comet ISON, after a long period of quiescence, may have foreshadowed the comet’s ultimate fate.

[7] Fragmentation here is referred to small pieces breaking away from the nucleus and not to the complete disruption/breakup of the nucleus.

Our global coma morphology campaign, which contains at least one set of images every day from November 1-17, provides an opportunity to explore this outburst timeframe in more detail. This task is challenging due to the non-uniformity of the data sets, but the temporal coverage is of value. During and after the reported increase in gas production on November 5, we see no changes in the dust coma morphology. However, the total gas production in this event increased by only ~50%, and it was noted that the dust production remained flat.

The November 12 outburst, on the other hand, showed an order of magnitude increase in both gas and dust productions, indicating a significantly more energetic event that is likely associated with the observed “wings”. Our campaign includes two sets of images, obtained at the appropriate time and with sufficient S/N to investigate this timeframe. Neither images acquired by Z.-Y. Lin and colleagues on November 13.85 nor images from Q.-Z. Ye and colleagues on November 13.99 show any identifiable “wing” structure, but images acquired by the same groups on November 14.86 and November 14.99, respectively, do show features similar to those reported by Boehnhardt et al. (2013). Based on this, we suggest that the faint “wings” reported by Boehnhardt et al. from images on November 14.2, first appeared in the coma sometime between November 13.99 and November 14.2 UT. In Figure 7, we show a sequence of images bracketing this time interval, showing the coma with no “wings”, the initial stage of their formation during November 14, and well developed “wings” lasting for at least three days.

The "wing" structure, as noted above signifies that a fragmentation event has occurred, and its timing corresponds to the (measured) “peak” in the production rates around November 14 (Opitom et al. 2013a, Combi et al. 2014), rather than to the start of the outburst on November 12. This indicates that the fragmentation is the result of the increased gas production, rather than the cause of the outburst. The morphology of the "wings" in both comet LINEAR and comet ISON, with nearly identical lobes in opposing directions, suggests that the source region is near the equator of a rapidly rotating nucleus, and the Earth is at low latitudes. These characteristics, combined with the timing of the events, allow us to propose a possible mechanism to explain these observations. As the nucleus approaches the Sun, heat penetrates into a pocket of volatiles, dramatically increasing the gas production. The higher gas drag, aided by erosion and centripetal acceleration near the equator, cause a piece of the surface to break off, revealing fresh ices. Material from this new active area, rotating with the nucleus, produces nearly axisymmetric lobes that appear as the "wings" seen in Figure 7, while contributing additional gas and dust to the peak of the outburst.

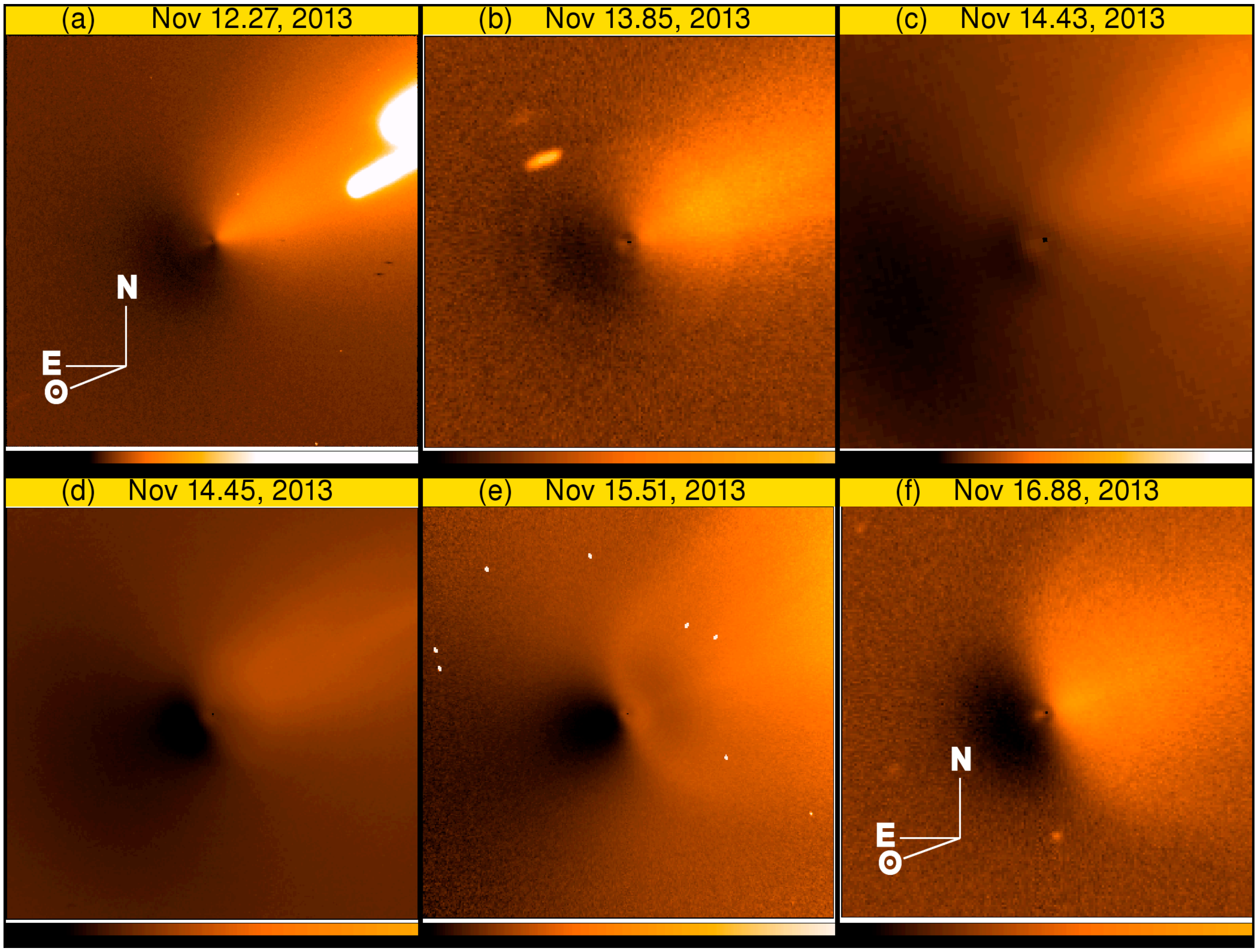

Figure 7. Images of comet ISON from November 12 ($r_h$~0.70 AU) to November 16 ($r_h$~0.56 AU) showing the development of the coma "wings" in comet ISON subsequent to the outburst of November 12. The time of observation is indicated at the top of each image and the time sequence is from left to right in the top row followed by left to right in the bottom row. The nucleus is at the center of each panel. Each panel is approximately 96,000 km across. The position angle of the Sun varies from 112° to 110° during this interval. All images were enhanced using the division by azimuthal median technique. The streaks present in some images are star trails. The observers were: (a) N. Howes and colleagues, (b) Z.-Y. Lin and colleagues, (c) D.C. Vogel, (d) C. Pruzenski and colleagues, (e) E. Gomez and colleagues, and (f) Z.-Y. Lin and colleagues.

By November 15 and 16, the "wing" structure was mature and well established. The dust in the "wings" further away from the nucleus was clearly affected by radiation pressure, being pushed significantly towards the anti-sunward direction. Unfortunately, we do not have sufficient observations to allow us to determine how long the "wings" persisted past November 16. Nor can we evaluate the effects of the weaker outburst reported by Opitom et al. (2013c) around November 18. They note the presence of “wings”, though it is not known if these are newly generated in the outburst or if they are simply the remaining vestiges of the November 14 features.

The equator being nearly edge-on around November 14 suggests that at that time the pole solution should lie close to the great circle defined by the sunward and anti-sunward directions as well as nearly in the skyplane, yielding a pole in the general direction of (RA=285°, Dec=-20°). Figure 5 shows the great circle solution for the sunward PA (dashed grey line) with a light grey envelope that denotes an uncertainty of 10° in the measured PA and an allowance for the pole being up to 30° out of the skyplane. Although this solution overlaps the HST solution discussed earlier, they are not required to be consistent because the pole could have changed due to the torques caused by outgassing in the intervening period between April and mid-November (cf. Samarasinha and Mueller 2013).

Many of our images from November 14 do not exhibit the “wings” seen in the high-resolution data. This is likely due to a combination of differences in astronomical seeing and S/N between images as well as the enhancement techniques being employed. However, the overlap of these data represents a means of connecting the signature from higher-resolution images to the lower-resolution images and for exploring the effects of enhancing those images. The data taken at the TRAPPIST Telescope at the La Silla Observatory in Chile around the same time as our observations show sharper “wings” if all images are enhanced with the same technique (C. Opitom; personal communications, 2015), which we attributed to the better seeing at the TRAPPIST location. As for enhancements, Boehnhardt et al. (2013) used Laplace filtering which is optimized for identifying spatial discontinuities. As a result, the strong sunward edge of the coma seen for images from November 14 in Figure 7, where we have enhanced the images by the more benign division of azimuthal median technique, may look like “wings” if enhanced with a Laplace filter (e.g., see Figure 6 in Farnham 2009). (For a detailed comparison of differences between enhancement techniques, the reader is directed to Samarasinha and Larson (2014).) A comparison of different images indicates that the same structure that shows well-developed “wings” in some observations, appears as a flattening or squashing of the sunward side coma in others (e.g., images acquired by D. Vogel on November 14.43 and C. Pruzenski and colleagues on November 14.45). This type of behavior can be used as a signature of a potential fragmentation event in future observations, when only low-resolution data are available.

An increase in polarization was observed for comet D/1999 S4 (LINEAR), which was tied to its fragmentation (Hadamcik and Levasseur-Regourd 2003). However, to our knowledge, no polarimetric observations are available from around mid-November until ISON’s demise although studies were conducted by comparing coma colors (e.g., Li et al. 2013) and polarization properties (e.g., Hines et al. 2014, Zubko et al. 2015) for comet ISON during the first half of 2013.

## 5. Summary and Conclusions

We collected many hundreds of images from nearly two dozen groups of amateur and professional observers, spanning nearly the entire time comet ISON was observable from the ground. During November 2013, when ISON was brightest but its visibility was severely restricted due to a small solar elongation, this allowed much better temporal coverage than would have been possible from a single location. For most of the apparition, comet ISON displayed no prominent morphological features. We attribute this partially to the relatively large geocentric distances to the comet. The main results based on this study are summarized below.

- A distinctive northwesterly sunward feature in the continuum marked the coma morphology during the pre-water dominated phase. This feature did not vary on diurnal timescales. We derive grain velocities of the order of 10 m $s^{-1}$ and the grains must have spent 2-4 weeks in the sunward side prior to merging with the dust tail.

- During the water-dominated phase, either the earlier continuum feature was absent or if it actually existed, it did not present itself prominently in the images. I.e., its skyplane-projected extent was of the order of the astronomical seeing or smaller and we provide an explanation for that based on the radiation pressure effects and the ground-based observational capabilities.

- Nearly two weeks prior to the perihelion, the comet started to display a variety of continuum features. The onset of this is related to the November 12 outburst reported by other authors. We constrain the “wing”-like features to have appeared between November 13.99 and 14.2 UT.

- This study shows that organized observing campaigns such as this one can collect observations from numerous observers, both professional and amateur, and assemble them into useful datasets. These campaigns may be most valuable in situations where any single observer can only obtain data during a small window of time, but contributions from many such observers provide coverage that leads to a more complete understanding of the spatial and temporal evolution of the comet.

## Acknowledgements

The NASA Planetary Astronomy Program with grant NNX14AG73G supported NHS and BEAM. MMK thanks NASA Planetary Astronomy grant NNX14AG81G. We thank the CIOC and many other individuals for providing publicity about this campaign. We thank C. Opitom for discussions related to their observations. MMK thanks Stephen Levine, Brian Skiff, and Larry Wasserman for help obtaining the data from Lowell Observatory. Z-YL thanks the staff of Lulin Observatory, Hsiang-Yao Hsiao, Chi-Sheng Lin, and Hung-Chin Lin, for their assistance with observations. We thank the two anonymous reviewers for their comments to improve the paper. This is PSI Contribution No. 629.